\documentclass[10pt,conference]{IEEEtran}
\pdfoutput=1
\usepackage{titlesec}

\titlespacing*{\subsection}
{0pt}{1.3ex plus 1ex minus .2ex}{1.0ex plus .2ex}

\ifCLASSINFOpdf
  \usepackage[pdftex]{graphicx}
   \graphicspath{{./}}
   \DeclareGraphicsExtensions{.pdf,.jpeg,.png}
\else
  \usepackage[dvips]{graphicx}
  \graphicspath{{../eps/}}
  \DeclareGraphicsExtensions{.eps}
\fi

\usepackage[utf8]{inputenc} 
\usepackage[T1]{fontenc}
\usepackage{microtype}
\usepackage{color, colortbl}

\newcolumntype{a}{>{\columncolor{Gray}}c}

\usepackage{amsmath, diagbox, hhline, booktabs}
\usepackage{gensymb}
\usepackage{nameref}
\usepackage{xspace}
\usepackage{todonotes}
\usepackage[font=small,labelfont=bf]{caption} 
\usepackage{soul}

\usepackage{array}
\newcolumntype{M}[1]{>{\centering\arraybackslash}m{#1}}

\usepackage{multirow} 
\usepackage{xcolor}

\usepackage{cite}
\usepackage[colorlinks,citecolor=blue,
            backref=section]{hyperref}
						
\makeatletter
\def\endthebibliography{%
  \def\@noitemerr{\@latex@warning{Empty `thebibliography' environment}}%
  \endlist
}
\makeatother

\usepackage{amsmath}
\usepackage{algorithmic}
\usepackage{array}

\ifCLASSOPTIONcompsoc
  \usepackage[caption=false,font=normalsize,labelfont=sf,textfont=sf]{subfig}
\else
  \usepackage[caption=false,font=footnotesize]{subfig}
\fi

\usepackage{fixltx2e}
\usepackage{stfloats}
\fnbelowfloat

\usepackage{multicol}
\usepackage{paralist}

\usepackage{url}
\usepackage{flushend}

\begin{document}
%margining the text
\sloppy
\title{{Towards an Approach to Pattern-based Domain-Specific Requirements Engineering}}

%this title is actual--------------------------
%two column title.
\author{\IEEEauthorblockN{Tatiana Chuprina, Daniel M\'endez$^{(1,2)}$, Vivek Nigam
 \IEEEauthorblockA{$^{(1)}$fortiss GmbH\\ 
 Munich, Germany\\
 $^{(2)}$Blekinge Institute of Technology\\
 Karlskrona, Sweden\\
% Email:\{chuprina,mendez,nigam\}@fortiss.org\\
}
 \and
 Marina Reich$^{(3,4)}$, Andreas	Schweiger$^{(3)}$\\
 \IEEEauthorblockA{
 $^{(3)}$Airbus Defence and Space GmbH\\
 Manching, Germany\\
 $^{(4)}$Technische University of Chemnitz\\
 Chemnitz, Germany\\
 %Email:\{marina.reich,andreas.schweiger\}@airbus.com
 \\}}}
%---------------------------------------------------
\maketitle
\begin{abstract}
Requirements specification patterns have received much attention as they promise to guide the structured specification of natural language requirements. By using them, the intention is to reduce quality problems related to requirements artifacts. Patterns may need to vary in their syntax (e.g. domain details/ parameter incorporation) and semantics according to the particularities of the application domain. However, pattern-based approaches, such as EARS, are designed domain-independently to facilitate their wide adoption across several domains.
Little is yet known about how to adopt the principle idea of pattern-based requirements engineering to cover domain-specificity in requirements engineering and, ideally, integrate requirements engineering activities into quality assurance tasks.
In this paper, we propose the \emph{Pattern-based Domain-specific Requirements Engineering Approach} for the specification of functional and performance requirements in a holistic manner. This approach emerges from an academia-industry collaboration and is our first attempt to frame an approach which allows for analyzing domain knowledge and incorporating it into the requirements engineering process enabling automated checks for requirements quality assurance and computer-aided support for system verification. 
Our contribution is two-fold: First, we present a solution to pattern-based domain-specific requirements engineering and its exemplary integration into quality assurance techniques. Second, we showcase a proof of concept using a tool implementation for the domain of flight controllers for Unmanned Aerial Vehicles. Both shall allow us to outline next steps in our research agenda and foster discussions in this direction.
\end{abstract}


\section{Introduction}
\label{Introduction}
The \emph{Naming the Pain in Requirements Engineering} initiative (NaPiRE)\footnote{See \url{www.napire.org}, last access 29/10/2019.} constitutes a large-scale survey research initiative to understand which practices industry practitioners use in requirements engineering (RE) and which problems they experience~\cite{Napire2016}. Based on recent results~\cite{Fernandez2018}, the most rated issues include “Underspecified Requirements” and “Weak Domain Knowledge”~\cite{Napire2016,Fernandez2018}. The latter rating strengthens our confidence in the value of an appropriate exploration of domain knowledge and its incorporation into the RE process as this is expected to improve the requirements' quality.

Pattern-based approaches, such as EARS~\cite{Mavin2009}, are well perceived in the RE community as a remedy for unstructured and imprecisely specified textual requirements. Moreover, such patterns are easy in application. 
Despite of all their merits, however, these patterns are domain-independent and an integration of domain specificity is not in their scope.

The \textit{main objective} of this research is to investigate \textit{how domain knowledge can be incorporated into the requirements specification for improving requirements engineering process}. In our current research efforts, we have focused on exploring how such patterns can improve the requirements' quality, while considering domain-specific aspects. Additionally, since RE is part of the overall systems development process and interconnected to all other stages (e.g., system design decisions are driven by requirements, system verification is related to valid requirements), we consider the impact of the domain knowledge not only on the RE process, but also on their propagation to the other development stages, such as system verification.

In this paper we summarize the related work (Section~\ref{RelatedWork}) and propose \textit{an Approach to Pattern-based Domain-Specific RE} to further close existing gaps. One particular challenge we focus on is the verification of requirements with a particular focus on performance requirements, i.e. requirements which describe the continuous input data space. 
The idea is to integrate defect-based testing for the verification of the system compliance according to performance requirements.
In Section~\ref{DSR}, we further contribute an example for the domain of flight controllers for Unmanned Aerial Vehicles (UAVs): we implemented first parts of our approach in the Model-Based Engineering tool AutoFOCUS3~\cite{Aravantinos2015} to show as a running example the Take-off Performance DSR; and we demonstrate, in Section~\ref{DefectsAndQA}, how domain knowledge can be incorporated into the requirements for integration into quality assurance techniques. In Section~\ref{DBT}, we then demonstrate how our approach supports system verification process, in particular, using defect-based testing techniques for system compliance of UAV controllers. We close with a conclusion along with a future work in Section~\ref{Conclusion}. 
\newpage

\section{Related Work}
\label{RelatedWork}
There exists a broad and elaborate body of knowledge for handling requirements quality issues and their causes as well as for improving or automating RE-related activities such as quality assurance or verification. Related work can be categorized into the following groups and will be discussed in the following paragraphs: (1) Natural language processing (NLP), (2) Pattern-based approaches including boilerplates, (3) Model-based approaches including domain-specific languages (DSL), and (4) Formal specifications.
 
\textbf{Natural language processing (NLP)}~\cite{Turing1950} aims at extracting the semantics of natural language for automated processing. Related techniques cope with the analysis of natural language available as text or speech. Those techniques applied are checking for, e.g., subjective language, vague pronouns or negative words. Requirements that contain such findings, are then considered to be potentially problematic. An introduction into using NLP for requirements quality assurance is summarized  in previvous work ~\cite{Femmer2016,Femmer:2014:RRC:2593812.2593817}. However, NLP allows for the check of a small subset of requirements quality properties only, such as unambiguities or consistencies in text. Exemplary valuable contributions are given by the authors in \cite{Chantree2006,Yang2011,FerrariGnesi2012,DBLP:conf/re/WinklerV16,Krner2009NaturalLS}. These approaches, however, do not cover yet the following aspects: (1) Aspects such as adequacy or pertinence cannot be tackled due to the focus on text, (2) NLP does not take into account domain knowledge provided in non-textual form such as figures, tables and symbols, and (3) NLP does not integrate the analyzed requirements into the verification of functional and performance requirements. Our approach intends to close those contemporary gaps.

\textbf{Pattern-based approaches} include controlled natural language (CNL) techniques. These approaches aim at controlling the requirements quality such as ambiguity or consistency. CNL operates with predefined rules to formalize and to structure textual requirements; for instance, boilerplates~\cite{books/daglib/0034874} are created based on frequent or appropriate patterns. EARS~\cite{Mavin2009} patterns and their extensions such as EARS-CTRL~\cite{Lucio2017EARSCTRLGC}, Parametrized Safety Requirements templates~\cite{Antonino2015ThePS}, and Adv-EARS~\cite{Majumdar2011} are prominent examples for providing templates to the engineer for gathering the requirements. Though some approaches propose to structure behavior requirements in tables, e.g., for defining a state automaton in a table view~\cite{Herrmannsdoerfer2008,Thyssen2013}, the following aspects are not in scope of existing pattern-based approaches: (1) Related approaches are applicable to functional requirements only, rather than to performance requirements, (2) they do not incorporate domain knowledge using figures, and (3) they do not cover requirements quality characteristics such as adequacy or pertinence. Also here we intend to close existing gaps with our contributions.

\textbf{Model-based approaches} aim at capturing requirements graphically, e.g., by applying UML/SysML~\cite{OMG2017,OMG2017a} or BPMN~\cite{BPMN}. The MIRA framework~\cite{Teufl2013a} is an example for a holistic model-based approach which considers quality assurance for functional requirements and an integration into a seamless development process. Model-based approaches also incorporate domain knowledge for a specific purpose~\cite{textDSL2015}. A domain model can support domain-specific modeling, as a domain model is capturing the concepts, terminology (defined by an ontology), scope, and features of the domain~\cite{Mernik}. This domain model can then be utilized for modeling e.g. a specific use case, requirement for a system or desired system behaviour in this domain~\cite{Stahl:2006}. We can include Domain-Specific Languages (DSL) for RE model-based approaches, and graphical (e.g., Arcadia/Capella~\cite{arcadia-capella}) or textual DSL for modeling requirements~\cite{textDSL2015}, because there is no strict distinction between them. Our DSRs elements also can be referred to DSL. However, when comparing existing approaches to our objectives, these approaches have several shortcomings: (1) Automated requirements quality assurance is not yet a prime objective of such approaches, (2) the integration of such approaches with defect-based testing is not supported, and (3) performance requirements are not taken into account. This is in scope of our contributions.

\textbf{Formal specification} offers precise logical definition (e.g., using temporal logic formulae), which copes with requirement quality issues such as ambiguity. Well-known techniques are Z~\cite{ISO13568} and B~\cite{abrial2005b} and their extensions Event-B~\cite{EventBAbrial} and Z++~\cite{ZplusLano}, respectively. Other approaches propose logical patterns by applying formal specifications, e.g., Specification Pattern System (SPS)~\cite{Filipovikj2014} or Real-time Specification Patterns~\cite{Konrad:2005}. However, formal specifications still suffer from the following limitations: (1) Requirements quality properties such as adequacy or pertinence are not covered, and (2) domain knowledge for RE is not exploited explicitly. Finally, also here we contribute to further closing existing gaps in literature.

\section{Approach}
\label{Approach} 
Inspired by our industry partner's project that aimed to analyze the reusable nature of UAV controller requirements, we developed \textit{an Approach to Pattern-based Domain-Specific RE} that can be used for requirements specification, requirements quality assurance and system compliance phases in a holistic manner. Moreover, our approach supports automation of the work-flow. The approach is based on the idea to collect descriptive data of the considered domain (domain knowledge) and incorporate them into graphical representation of requirements (in short, a picture with appended meta-data), so that such approach can control quality of the requirements and support integration the RE with a system verification process (here with defect-based testing techniques).

We analyzed the given requirements for quadcopter controllers which are typically used by control engineers (CEers). We discovered that CEers use \textit{graphical representations} of the requirements along with textual specifications, because presenting requirements graphically is more informative way for explaining system design, than describing it with sentences of text. Indeed, as we describe in Section~\ref{DefectsAndQA}, the domain knowledge can be intuitively represented by a simple picture of the desired performance of the quadcopter controller. 

Because of requirements repetition from project to project for the same kind of system (e.g. a basic drone should perform take-off or landing maneuvers), the graphical representation of the requirements can be considered as graphical patterns. Comparing to EARS~\cite{Mavin2009} domain-independent textual patterns, we see an advantage of our approach in graphical pattern-based presentation of requirements with domain knowledge incorporation. These patterns were recently introduced in a short paper~\cite{DBLP:conf/se/ReichCN19} and called \textit{Domain-Specific Requirements} (DSRs).

\begin{figure}[t]
\centering
\setlength{\belowcaptionskip}{-5pt}
\includegraphics[width=0.47\textwidth]{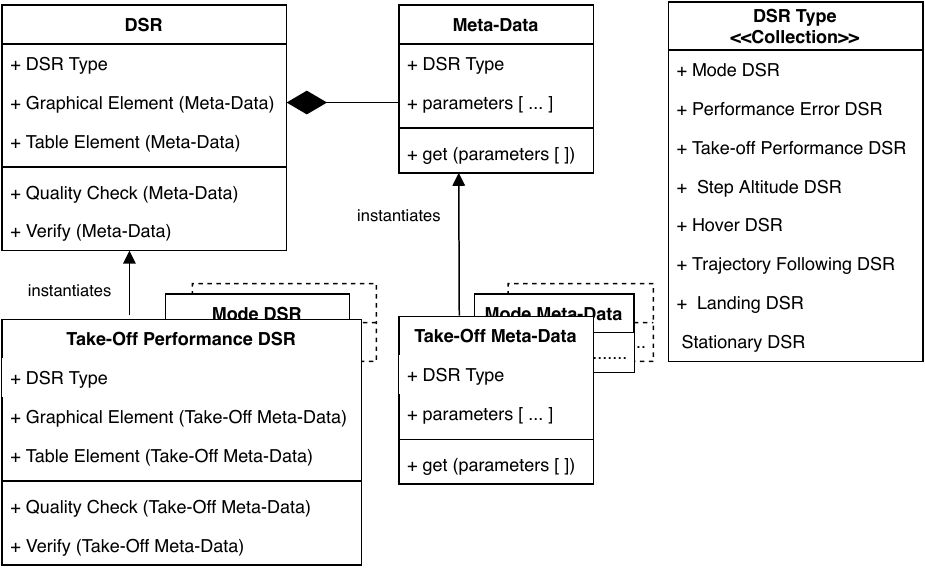}
\caption{an Approach to Pattern-based Domain-Specific RE}
\label{fig:Process}
\vspace*{-3mm} 
\end{figure}

We identified that DSRs for UAV Controller include eight patterns: (1) Mode DSR, specifying the UAV modes and transitions; (2) Performance Error DSR, specifying the performance error when transiting from one mode to another; (3) Take-off DSR and (4) Step Altitude DSR, specifying the performance of take-off UAV maneuvers; (5) Hover DSR, (6) Trajectory Following DSR, (7) Landing DSR, (8) Stationary DSR, these four DSRs as well as Take-Off DSR describe performance modes of the UAV. Figure~\ref{fig:Process} presents meta-model of DSR collection for UAV Controller we propose in this paper. DSR-elements are model elements such as, figures and tables, resembling usual (textual) requirements of the domain.

Such pattern representations are known to be informative and serve for better understanding, explanation of requirements \cite{Storrle:2017:CMU:3084226.3084256} improving the communication process between stakeholders within system life-cycle. However, differently from usual textual requirements containing figures, the (meta-)data in DSRs have precise meaning, thus enabling semi-automated quality checks. 
Figures help visualize the key aspects of a requirement, such as, variables and parameters. Tables are then used to specify the properties of these aspects, such as, variable ranges. Further in this paper, we present example with take-off maneuver of UAV to demonstrate this. Figure~\ref{fig:TakeoffDSR} shows "Take-off DSR" with the graphical element of the take-off maneuver, table elements for structuring the requirement and domain-specific parameters, which describes possible scenarios of the take-off maneuver of UAV.

\section{Take-Off Performance DSR}
\label{DSR} 
\begin{figure}[t]
\centering
\setlength{\belowcaptionskip}{-15pt}
\includegraphics[width=0.45\textwidth]{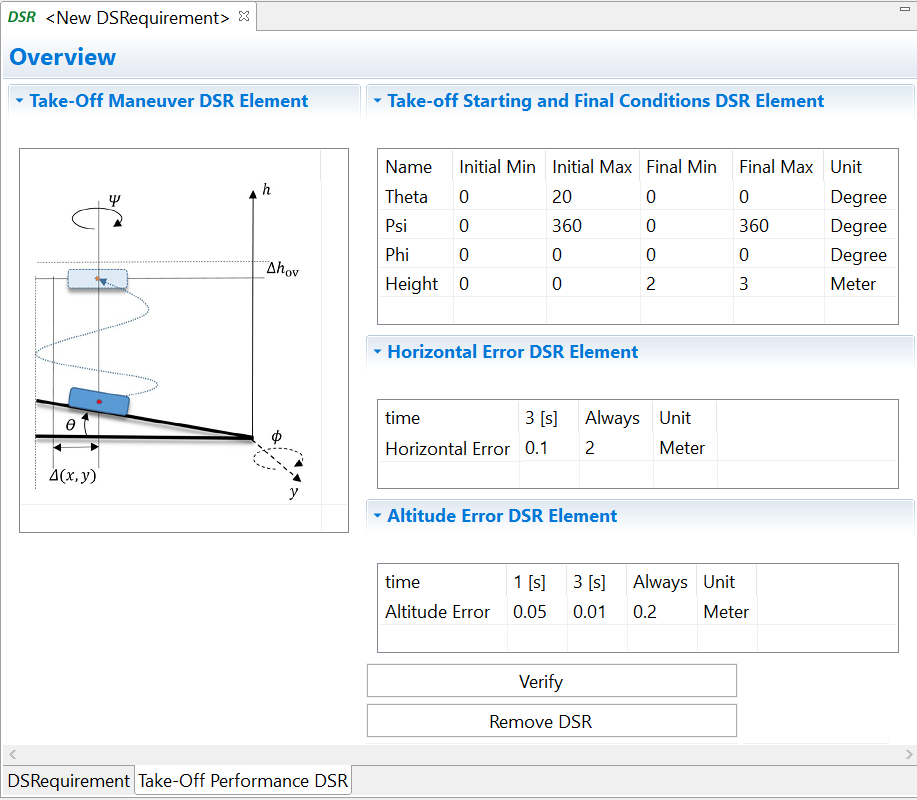}
\caption{Take-Off Performance DSR Prototype in AutoFOCUS3.}
\label{fig:TakeoffDSR}
\vspace{-2mm}
\end{figure}
We illustrate DSRs with a Take-Off performance requirement for UAV controllers. The Take-Off Performance DSR specifies the control performance of the UAV while carrying out a take-off maneuver. Its elements use the figure and the tables presented on Figure~\ref{fig:TakeoffDSR}. The idea is that the take-off maneuver illustrates graphically the key parameters relevant for this requirement. Then the exact bounds for these values are established by using the remaining DSR elements. The values in the tables (DSR elements) are not to be taken too seriously, but serve only for illustrative purpose. Take-Off Performance DSR is composed by four DSR elements:

\textbf{Take-Off Maneuver DSR Element:} This element is a figure illustrating the main maneuver parameters. An instance of this DSR element is depicted on the left side of Figure~\ref{fig:TakeoffDSR}. It is a commonly used illustration by control engineers to describe the takes-off movement of UAV  and the performance criteria. Therefore, it naturally incorporating domain knowledge.
The take-off maneuver consists of: taking off from an initial position point; specified
rotation angles about the axis $(x,y,z)$ of the UAV: \emph{pitch} ($\Theta$), \emph{yaw} ($\Psi$) and \emph{roll} ($\Phi$); moving upwards to the designated height $h$; two UAV performance variables: the overshoot ($\Delta h_{ov}$) and the horizontal deviation ($\Delta(x,y)$).

\textbf{Take-Off Starting and Final Conditions DSR Element:} The second DSR element is a table for specifying the ranges of the UAV parameters ($\Theta, \Psi, \Phi$ and $h$) in initial and final position points. The values in the Figure~\ref{fig:TakeoffDSR} specify, for example, that the UAV can take-off to a height between 2 or 3 meters when starting from any surface with inclination ($\Theta$) of at most 20 degrees.

\textbf{Horizontal Deviation DSR Element} is a table for specifying the performance of the UAV take-off maneuver in terms of the maximal horizontal deviation ($\Delta(x,y)$) (depicted as Horizontal Error on Figure~\ref{fig:TakeoffDSR}) in any direction from the take-off point at times \emph{after starting the take-off maneuver}.
A special entry \emph{always} indicates the global property that shall be satisfied at all times by the take-off maneuver. For example, at all times the horizontal deviation shall be not greater than 2 meters. Also, in the table a particular \textit{time steps} of take-off maneuver can be specified, e.g., 3 seconds [s] after the system became airborne, the horizontal deviation shall not be greater than 0.1 meter.

 \textbf{Altitude Deviation DSR Element} is a table for specifying the performance of the UAV take-off maneuver in terms of the maximum admissible altitude error ($\Delta h_{ov}$) at different times \emph{after reaching the target height.} As with the Horizontal Deviation DSR Element, the table also contains the entry \emph{always}. 

Notice that the time steps for the Horizontal and Altitude Deviation DSRs may differ and not be limited in number.  

Organizing DSRs in this way has some advantages. It incorporates naturally the domain-specific terminology by means of figures and tables as opposed to more sophisticated machinery, such as formal specifications. Moreover, such a DSR enables the semi-automated methods for requirement quality assurance and system compliance, as the meaning of the values can be inferred from the meta-data (domain-specific parameters). The explanation goes in Sections~\ref{DefectsAndQA} and~\ref{DBT}.

\subsection{Domain-Specific Defects} 
\label{DefectsAndQA}
This Section proposes a set of  Domain-Specific Defects (DSDs) for Take-off Performance DSR described in Section~\ref{DSR}. The data in {\color{red}red} illustrate Domain-Specific Defects presented as Table~\ref{defects}.

In our research we took the requirement quality characteristics proposed in~\cite{vanLamsweerde:2009} as a base set for DSR quality assurance.

We first argue that the DSRs do not suffer from the defects \textit{Unintelligibility}, \textit{Unstructured Textual Requirements}, \textit{Invisible Dependencies Between Items}. Below we present a list of the Domain-specific defects and respective to these defects quality characteristics (QC) in format \textbf{[defect-QC]}.
\begin{compactitem}
\item \textbf{Unintelligibility - Measurability:} defect indicates that a requirement is bad-defined and, consequently, not prepared for the evaluation process.
The use of domain knowledge in DSR, incorporated into the DSR elements support the precision necessary for evaluation. For example, the Take-off Performance DSR uses well-established terms for control engineers, such as pitch,roll and yaw angles. 
\item \textbf{Unstructured Textual Requirements - Good Structuring:} defect indicates the problem of a bad-structured textual requirement, where the structural links among DSR elements are not highlighted. The DSR elements in Take-Off Performance DSRs proposed have a logical connection. For example, the \emph{Take-Off Maneuver DSR element} illustrates the parameters used by the \emph{Take-Off Starting and Final Conditions DSR Element} and \emph{Altitude Error DSR Element}.
\definecolor{Gray}{gray}{0.9}
\begin{table}[bp]
\caption{Example of Domain-Specific Defects for Take-Off Starting and Final Conditions DSR Element.}
\begin{center}\begin{small}
\begin{tabular}{|a|m{1cm}|m{0.8cm}|m{0.8cm}|m{0.8cm}|m{1cm}|} 
\hline
&\multicolumn{2}{c|}{\textbf{Initial}}&\multicolumn{2}{c|}{\textbf{Final}}& \multirow{2}{4em}{\textbf{Unit}} \\
& min & max & min & max &  \\ 
\hline 
Theta ($\Theta$) & 0 & $\emph{\color{red}\textbf{95}}$ & 0 & 0 & Degree\\ \hline 
Psi ($\Psi$) & $\emph{\color{red}\textbf{320-360}}$ & 360 & 0 & 360 & Degree \\ \hline
Phi ($\Phi$) & $\emph{\color{red}\textbf{0.001}}$ & $\emph{\color{red}\textbf{--}}$ & 0 & 0 & $\emph{\color{red}\textbf{--}}$\\ \hline
Height (${h}$) & $\emph{\color{red}\textbf{--}}$ & $\emph{\color{red}\textbf{--}}$ & $\emph{\color{red}\textbf{-1}}$ & 2 & Meter \\ 
\hline
\end{tabular}
\label{defects}
%\vspace*{-2mm}
\end{small}\end{center}
\end{table}

\item \textbf{Invisible Dependencies Between Items - Traceability:} The same argument can be used for this defect. The dependencies are explicit in the construction of DSRs, through the incorporation of domain knowledge within requirements (DSR-Elements).
\item \textbf{Omission - Completeness:} 
The Take-Off Performance DSR suffers from Omission if it does not contain the data in its Table DSR Elements. For example, missing values for the initial conditions of the Take-Off maneuver in the table Take-Off Performance DSR Element, however the domain-specific parameters for these conditions are presented at a pictured Take-Off Maneuver DSR Element \textit{(Psi, Theta, Phi, $\Delta{(x,y)}$ horizontal or $\Delta{h\_ov}$ vertical deviations)}.
\item \textbf{Imprecise - Unambiguity:} 
The Take-Off Performance DSR provides a template-based table view, which structures the given data of requirements into the certain format: \textit{[Parameter] = [Value]~[Unit]}. This helps to avoid \textit{Imprecise} information defect, such as parameters are declared in an imprecise way: \textit{initial $\Psi$ is from 0 to 320 or 360, or without unit:} \textit{$\psi$=$[$0;320-360$]$ $[$-$]$}.
\item \textbf{Contradiction - Consistency:}  
If every parameter and its unit are not defined consistently w.r.t UAV domain and related DSR Elements, then we can conclude that DSR defected,e.g., $\Phi$=0,001[-] and $\Theta$=[0;20][degree].
\item \textbf{Incorrect Domain Laws Description - Adequacy:} The incorporation of domain-specific knowledge in DSRs is valuable in defining precisely these defects. For the Take-Off Performance DSR, it is defective if one specifies unrealistic parameters without rationale. According to domain rules, UAV cannot take-off from inclination $\Theta$ = 95 degree or take-off attitude cannot be negative. 
\item \textbf{Superfluous - Pertinence:}
Take-Off Performance DSR will be defective if it specifies variables which are out of its domain-specific set, e.g., \textit{landing point(x,y,h)} = (0,0,0), which is not a relevant parameter for taking-off UAV maneuver.
\end{compactitem}
We apply static analysis methods for requirements quality checks comparing predefined domain-specific data baseline with the data incorporated in Take-Off Performance DSR. This comparison can be performed by setting predefined constraints for data in DSRs, e.g., \textbf{constraint:} \textit{"allowed range for inclination parameter $\Theta$ $\in$ [0;20]"}. Such approach was proposed by~\cite{VinceConstraints} and allows automation of the quality checks.

\subsection{DSRs Integration with Defect-based Testing}
\label{DBT}
The verification of performance requirements, such as, for UAV controllers described above, is challenging because there are infinitely many test cases to consider (instead of precise values, the take-off parameters (i.e.$\Theta$, $\Psi$, $h$) are the allowed ranges). 

To address this challenge, defect-based testing (DBT) approaches have been successfully deployed~\cite{Pretschner2017}. 

The overall DBT approach is to build test-cases so that to maximize the chance of triggering a defect. For example, when testing the logic of a software, one tests its corner cases, e.g., by applying Boundary Value Analysis~\cite{BVA}.

In this research, we review the definition of the DBT artifacts and explain how these are instantiated for our approach continuing with the example of Take-Off Performance DSR:
\begin{compactitem} 
\item \textbf{Specification:} The specification represents properties that shall be tested. These have to be verified for the implementation. Our DSRs contain these specifications by design. For example, the Take-Off Performance DSR contains the initial conditions for the take-off maneuver and maximum horizontal deviation at a given elapsed time. 
\item \textbf{Implementation Model:} The implementation represents the system under test. It is an executable software design of the system under development. For UAV controller development, the implementation model is developed in Matlab/ Simulink. This model is assumed to reflect as close as possible the actual UAV to be developed;
\item \textbf{Environment Model:} If the system interacts with its environment, the environment model can be provided. For UAV controller development, the implementation model incorporates the environment model, namely, in the flight dynamics model. The environment, Quadcopter Dynamic Modeling and Simulation package (Quad-Sim)\footnote{git repository:\url{https://github.com/dch33/Quad-Sim}}\footnote{\url{https://www.mathworks.com/matlabcentral/fileexchange/48053-quad-sim}} was used for our simulation; 
\item \textbf{Failure Model:} A failure is the observed difference between actual and intended behavior. This model is used to determine which test-cases have a greater chance to find a defect in the actual UAV implementation (and not the simulation model). For the Take-Off Maneuver DSR, the failure model is how close the UAV does not comply with the altitude and horizontal errors. 
\end{compactitem}

We now put these components together to generate test-cases using DBT methodology on the Take-Off Performance DSR:

\textbf{(1) Input Space Partitioning:} We partition the input space, namely, the values for the initial yaw($\Psi$), pitch($\Theta$) (we elide $\Phi$ for simplicity) and intended height($h$), specified in the Take-Off Performance DSR: 
\vspace*{-1mm}
  \[
\begin{array}{lll}
  \Theta \in [0,6] & \Theta \in~ ]6,12] & \Theta \in~ ]12,20]\\
  \Psi \in [0,120]  &  \Psi \in~ ]120,240]  &  \Psi \in~ ]240,360]\\
   h \in [0,1] & h \in~ ]1,2] & h \in~ ]2,3]\\
\end{array}    
  \]
\vspace*{-3mm}

\textbf{(2) Generate Random Take-Off Scenarios:} For each partition $P_j$, we generate $n$ random take-off scenarios, $S_1,\ldots, S_n$, by randomly generating values for $\Theta, \Psi, h$ that belong to the partition,e.g. $\Theta = 7.4, \Psi = 101, h = 2.3m$. 

In our use case we specified 2 test-cases for every partition.
The greater the number of scenarios generated, the greater is the effort and the greater are the chances for finding a defect. 

Note that the number of take-off scenarios($n$) and the granularity of the partitions depend on the available computational power and the desired level of precision.

\textbf{(3) Run Take-Off Simulations:} Using the implementation/environment model, we carry out one simulation for each generated take-off scenario, $S_i$. The output is a simulated take-off maneuver, $\Xi_i$. 

\textbf{(4) Compute Failure Model:} 

For our example, we generate test-cases that maximize the chance of finding defects on the admissible horizontal deviation. (For the altitude deviation is similar.) 

$\Delta(x,y)$ is the horizontal deviation in the Take-Off Performance DSR. For each simulated take-off maneuver, $\Xi_i$ for the scenario $S_i$, we compute the maximum horizontal deviation, $max({\Xi_i})$. Afterwards, we compute for $S_i$ its \emph{defect-value} as follows:
\vspace*{-2mm}
\[
dfb(S_i) =
\left\{
\begin{array}{ll}
  \infty & \textrm{if $max({\Xi_i}) > \Delta(x,y)$}\\
  \frac{1}{\Delta(x,y) - max({\Xi_i})} & \textrm{otherwise}
\end{array}\right.
  \]  
\noindent Intuitively, the greater the defect-value, the greater is the chance for the scenario $S_i$ that generated $\Xi$ to trigger a defect for the actual UAV.  

\textbf{(5) Consolidate Defect-Values for Partitions:} Let $dfb(S_1), \ldots, dfb(S_n)$ be the defect-values of the scenarios generated for the partition $P_j$ in step (2). The defect-value of the partition $P_j$, $dfb(P_j)$  is equal to $dfb(S_1) + \cdots + dfb(S_n)$. 

\begin{figure}[tp]
	\centering
	\setlength{\belowcaptionskip}{-5pt}
		\includegraphics[width=0.35\textwidth]{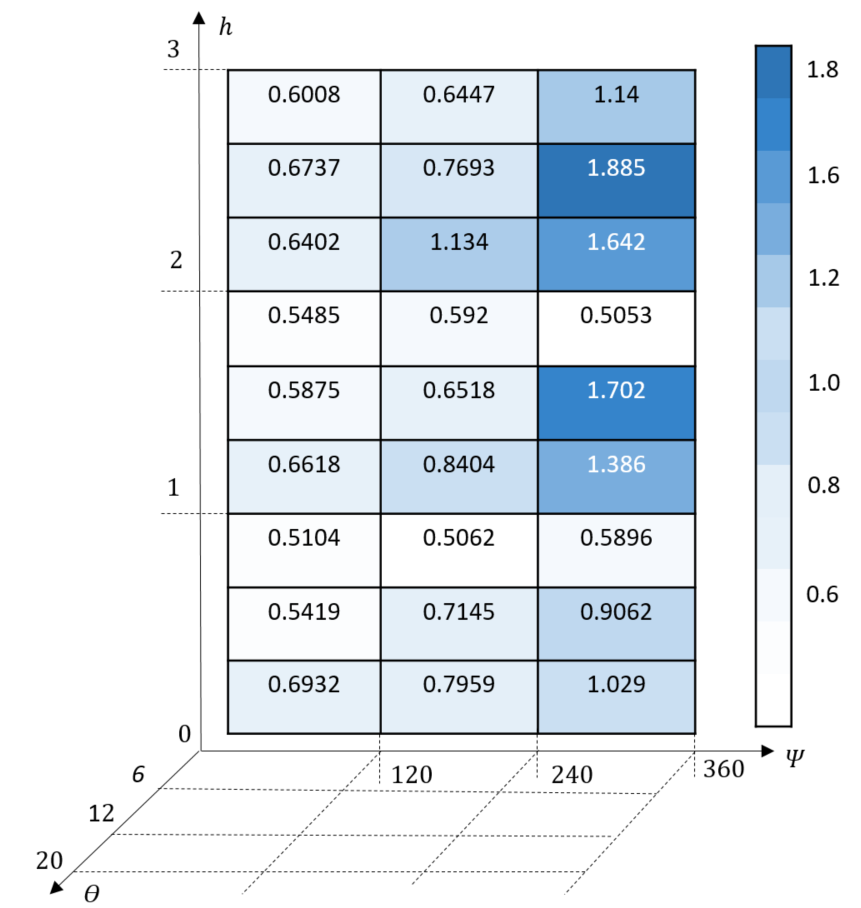}
	\caption{Illustration of Defect-Values for Partitions.}
	\label{fig:Sim_output}
	\vspace{-2mm}
\end{figure}

Figure~\ref{fig:Sim_output} is an illustration of an output of the procedure $(1) - (5)$ as a heat map, where defect-based points inverted to by 1 and presented in color (the darker color, the more chance of triggering the defect). Such output presentation supports prioritization of the test-case scenarios, and indicates the partition-candidate for more precise testing.

\section{Conclusion and Future Work}
\label{Conclusion}
In this paper we present our first attempts towards \textit{an Approach to Pattern-based Domain-Specific RE} illustrating the concept for performance requirements with Take-off DSR. Using this example, we demonstrate how our holistic approach integrates domain-specific requirements specification, computer-aided quality assurance techniques and system verification techniques, such as defect-based testing. Providing Take-off performance DSR, we explain how domain knowledge 
can be incorporated in performance requirements specification to improve quality of requirements and to support system verification process. 

A further most interesting and challenging question is related to non-functional requirements, i.e., how our approach can be applied to requirements specifications which describe such system aspects as safety and security. 

Furthermore, our initial results (as well as those of upcoming) require an evaluation at practice. Therefore, the next stage of the research can be an empirical study for validating our approach and publishing the achieved results. Finally, the approach requires an investigation on scalability by applying it for other domains, e.g., self-driving cars (AI-based systems domain).

Therefore, we believe that our \textit{Approach to Pattern-based Domain-Specific RE} can be helpful and poses actual and interesting questions for the community.
\smallskip
\textbf{\textit{Acknowledgements.}} 
Our research is a part of the global research funded by DAAD, in cooperation with Institute of Flight System Dynamics (FSD) at Technical University of Munich (TUM), Germany and Federal University of Paraíba, Brasil. Also we thank PhD student Daniel Morais for the support in this research.

% ---- Bibliography ----
 \bibliographystyle{abbrv}
 \bibliography{dsre}
\end{document}